# SPEECH ENHANCEMENT USING PITCH DETECTION APPROACH FOR NOISY ENVIRONMENT

RASHMI MAKHIJANI

Department of CSE, G. H. R.C .E., Near CRPF Campus,Hingna Road,
Nagpur , Maharashtra, India
rashmi.makhijani2002@gmail.com

URMILA SHRAWANKAR

Department of CSE, G. H. R.C .E., Near CRPF Campus,Hingna Road,
Nagpur , Maharashtra, India
urmila@ieee.org

Dr. V. M. THAKARE

Department of CSE, S. G. B.,Amravati University,
Amravati , Maharashtra, India

Abstract :
Acoustical mismatch among training and testing phases degrades outstandingly speech recognition results. This problem has limited the development of real-world nonspecific applications, as testing conditions are highly variant or even unpredictable during the training process. Therefore the background noise has to be removed from the noisy speech signal to increase the signal intelligibility and to reduce the listener fatigue. Enhancement techniques applied, as pre-processing stages; to the systems remarkably improve recognition results. In this paper, a novel approach is used to enhance the perceived quality of the speech signal when the additive noise cannot be directly controlled. Instead of controlling the background noise, we propose to reinforce the speech signal so that it can be heard more clearly in noisy environments.The subjective evaluation shows that the proposed method improves perceptual quality of speech in various noisy environments. As in some cases speaking may be more convenient than typing, even for rapid typists: many mathematical symbols are missing from the keyboard but can be easily spoken and recognized. Therefore, the proposed system can be used in an application designed for mathematical symbol recognition (especially symbols not available on the keyboard) in schools.

1. Introduction

The perceptual quality of speech, which is defined as the overall quality of perception measured in terms of intelligibility, clarity, and naturalness, is seriously degraded by ambient noise. Many methods for improving the perceptual quality of speech in a noisy environment have been proposed and are applied. Each method suggests the enhancement of perception-related speech features such as signal-to-noise ratio (SNR), loudness, and high-band components [1]–[4].
This paper proposes a new method is designed to solve the problem of unpredictable performance of conventional methods. The proposed method will be applied for recognizing mathematical symbols[5] in a school environment.

## 2. Methodology

Speech recognition is, in its most general form, a conversion from an acoustic waveform to a written equivalent of the message information. To process the speech signal digitally, it is necessary to make the analog waveform discrete in both time (sample) and amplitude (quantize). Speech recognition process consists of various steps such as Speech Acquisition, Speech Preprocessing, Feature Extraction, Training and Recognition. The main stress in this paper is on the second step i.e. Speech Preprocessing.

During the first step i.e. speech acquisition, speech samples are obtained from the speaker in real time and stored in memory for preprocessing. Pre-processing is a critical process performed on speech input in order to develop a robust and efficient system [6]. It is mainly performed in a few stages such as A/D conversion, End point, Pre-emphasis and speech enhancement. The first stage is the Analog-to-Digital (A/D) conversion where the analog speech signal is converted into a digital signal. The second stage is the removal of silent segment from the captured speech signal, otherwise known as end-point detection. Endpoint detection refers to the removal of silence portion of the speech data. The two most widely used end-point detection methods in use are the short-time energy based method (STE) and the zero-crossing method (ZCR).STE will be method implemented for this process in this project. Basically, the speech signal will be divided into 0.5ms frames and compared with the average energy of the speech signal. Frames with energy below the threshold set will be discarded. Retained frames will be combined to form the final speech data for further speech processing.

The third stage in preprocessing is Pre-emphasis which is used to enhance the high frequencies of speech signal. There are two important factors for doing this:
(1) To enhance the specific information in the higher frequencies of speech.
(2) To negate the effect of energy decrease in higher frequencies in order to enable proper analysis on the whole spectrum of the speech signal.

After pre-emphasis stage, speech enhancement techniques [7] are used based on Pitch detection .The pitch determination is very important for many speech processing algorithms. For example, the concatenative speech synthesis methods require pitch tracking on the desired speech segments if prosody modification is to be done. Pitch is also crucial for prosodic variation in text-to-speech systems and spoken language systems. In this paper, pitch detection methods using cepstrum method is used.

## 3. Pitch Detection via Cepstral Method

Cepstral analysis provides a way for the estimation of pitch. If we assume that a sequence of voiced speech is the result of convoluting the glottal excitation sequence e[n] with the vocal tract's discrete impulse response _[n]. In frequency domain, the convolution relationship becomes a multiplication relationship. Then, using property of log function log AB = log A + log B, the multiplication relationship can be transformed into an additive relationship. Finally, the real cepstrum of a signal
$s[n] = e[n] * \theta[n]$ is defined as

$$c[n] = \frac{1}{2\pi} \int_{-\pi}^{\pi} log|S(w)| e^{jnw} dw \quad (1)$$

where

$$S(w) = \sum_{n=-\inf}^{\inf} s[n] e^{-jwn}. \quad (2)$$

That is, the cepstrum is a Fourier analysis of the logarithmic amplitude spectrum of the signal. If the log amplitude spectrum contains many regularly spaced harmonics, then the Fourier analysis of the spectrum will show a peak corresponding to the spacing between the harmonics: i.e. the fundamental frequency. Effectively we are treating the signal spectrum as another signal, then looking for periodicity in the spectrum itself.

The cepstrum is so-called because it turns the spectrum inside-out. The x-axis of the cepstrum has units of quefrency, and peaks in the cepstrum (which relate to periodicities in the spectrum) are called rahmonics. To obtain an estimate of the fundamental frequency from the cepstrum we look for a peak in the quefrency region corresponding to typical speech fundamental frequencies (1/quefrency).

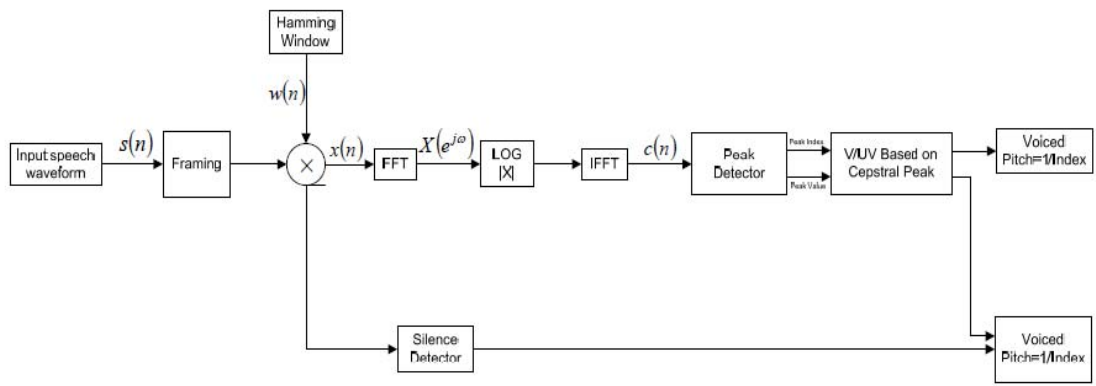

Fig. 1 Flow Graph of Cepstrum analysis

Cepstral analysis separates the effects of the vocal source and vocal tract filter. speech signal can be modeled as the convolution of the source excitation and vocal tract filter, and a cepstral analysis performs deconvolution of these two components. The high-time portion of the cepstrum contains a peak value at the pitch period. Figure 1 shows a flow diagram of the cepstral pitch detection algorithm.The cepstrum of each hamming windowed block is computed. The peak cepstral value and its location are determined in the frequency range of 60 to 500 Hz as defined in the autocorrelation algorithm, and if the value of this peak exceeds a fixed threshold, the section is classified as voiced and the pitch period is the location of the peak. If the peak does not exceed the threshold, a zero-crossing count is made on the block. If the zero-crossing count exceeds a given threshold, the window is classified as unvoiced.Unlike autocorrelation pitch detection algorithm which uses a low-passed speech signal, cepstral pitch detection uses the full-band speech signal for processing. Figure 2 shows the result of applying cepstrum method of pitch detection on the sample wav file.

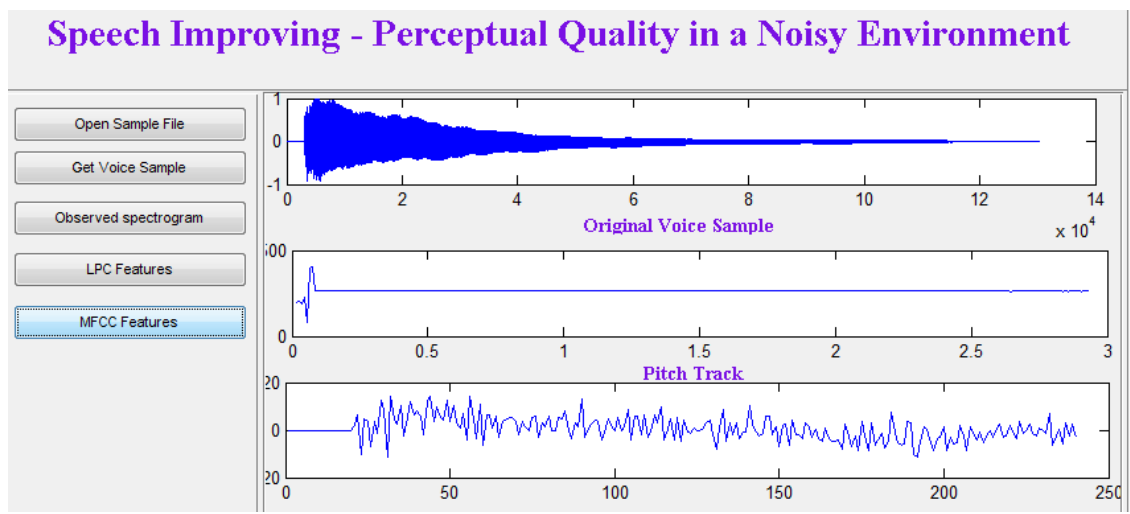

Fig. 2 Result of Cepstrum analysis on speech signal

4. **Voice features extraction**

Voice feature extraction, otherwise known as front end processing is performed in both recognition and training mode. Feature extraction converts digital speech signal into sets of numerical descriptors called feature vectors that contain key characteristics of the speech signal. Evaluation of the different types of feature extracted from voice to determine their suitability for recognition is the third step for this paper. The current most popular and widely known features used are the Linear Prediction Coefficients (LPC), Linear Prediction Cepstral

Coefficients (LPCC) & Mel-Frequency Cepstral Coefficients (MFCC)[8]. As such, in this paper the MFCC is used for feature extraction.

**4.1.** *Mel-Frequency Cepstral Coefficients (MFCC)*

Mel-frequency Cepstral coefficient is one of the most prevalent and popular method used in the field of voice feature extraction. The difference between the MFC and cepstral analysis is that the MFC maps frequency components using a Mel scale modeled based on the human ear perception of sound instead of a linear scale [9]. The Mel-frequency cepstrum represents the short-term power spectrum of a sound using a linear cosine transform of the log power spectrum of a Mel scale. The formula for the Mel scale is

$$M = 2595 \log_{10}\left(\frac{f}{700} + 1\right) \qquad (3)$$

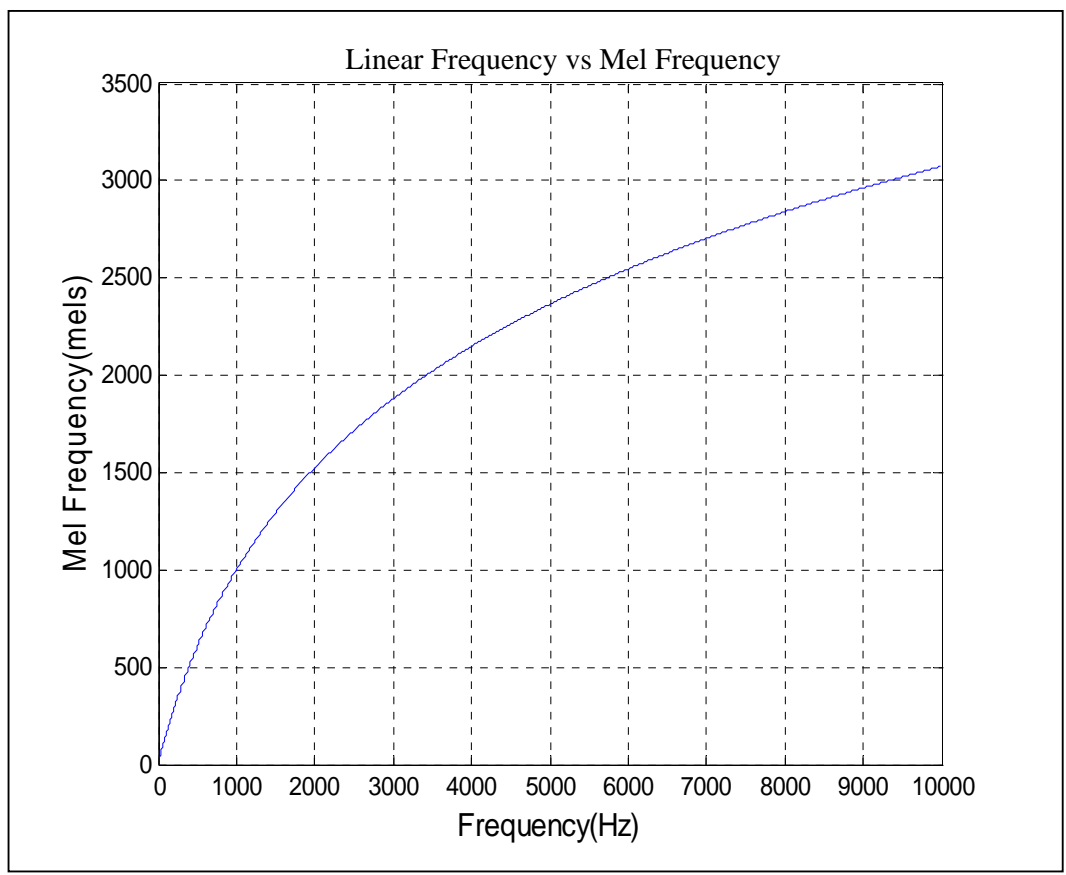

Fig. 3 Mel Scale plot

Vergin [10] mentioned that MFCC as frequency domain parameters are much more consistent and accurate than time domain features. Vergin [10] listed the steps leading to extraction of MFCCs: Fast Fourier Transform, filtering and cosine transform of the log energy vector. According to Vergin [11], MFCCs can be obtained by the mapping of an acoustic frequency to a perceptual frequency scale called the Mel scale. MFCCs are computed by taking the windowed frame of the speech signal, putting it through a Fast Fourier Transform (FFT) to obtain certain parameters and finally undergoing Mel-scale warping to retrieve feature vectors that represents useful logarithmically compressed amplitude and simplified frequency information. Seddik [12] mentioned that MFCC are computed by applying discrete cosine transform to the log of the Mel-filter bank. The results are features that describe the spectral shape of the signal. Rashidul describe the main steps for extraction of MFCC, shown on figure 4. The main steps are as follow: pre-emphasis, framing, windowing, perform Fourier fast transform FFT), Mel frequency warping, filter bank, logarithm, discrete Cosine transform (DCT).

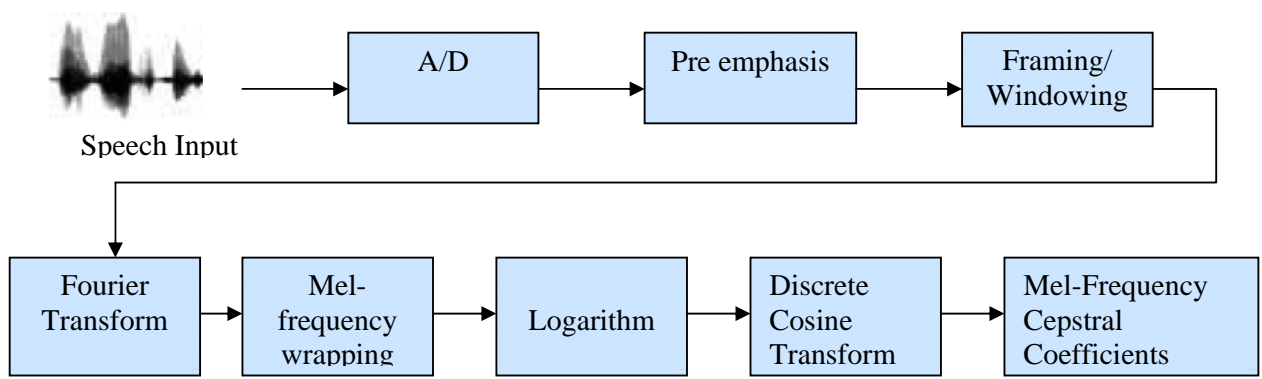

Figure 4 Block diagram of Mel-Frequency Cepstral Coefficient

MFCC uses banks of filters to wrap the frequency spectrum onto the Mel-scale that is similar to how the human ear perceives sound. The filters of the Mel-scale are linear at low frequencies but logarithm at high frequencies to imitate the human hearing perception. For this project, the filters of the mfcc will be adapted from *Voicebox: Speech Processing Toolbox for MATLAB by Mike Brooks*. In this paper, MFCC are extracted by passing the frames of the windowed speech signal into the mfcc.m function written. Figure 5 shows the result of applying LPCC & MFCC method on the speech signal.

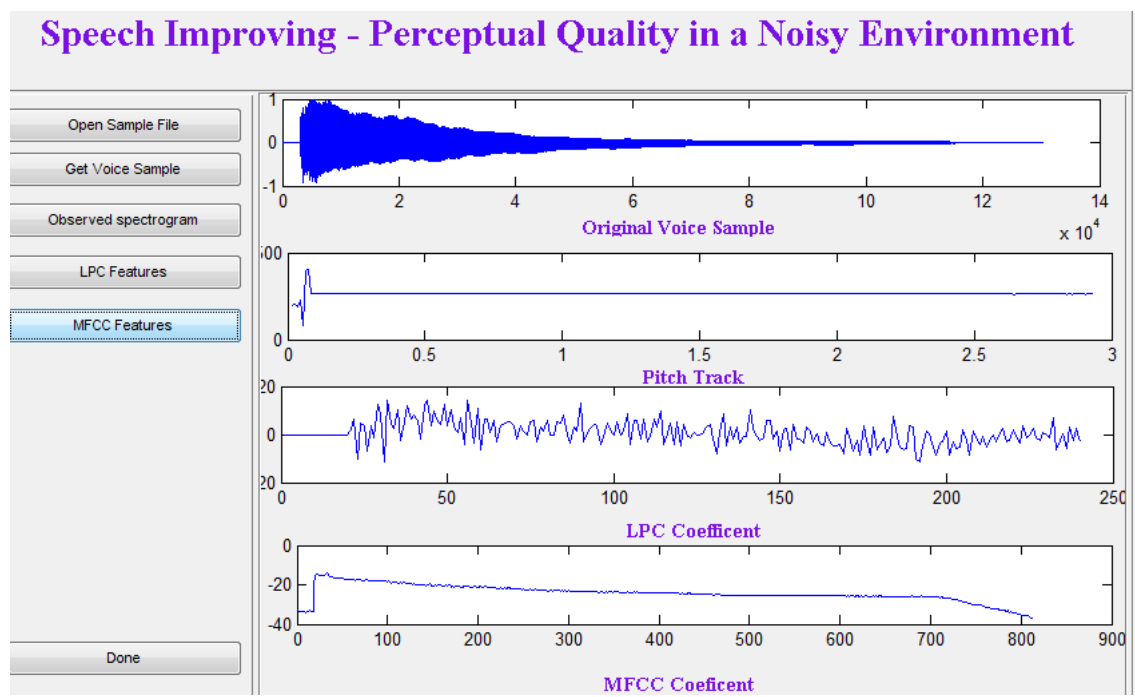

Figure 5 Result of applying LPCC & MFCC method of Feature Extraction on speech signal.

The main advantage of MFCC is the robustness towards noise and spectral estimation errors under various conditions [13]. A. Reynolds did a study on the comparison of different features and found that the MFCC provides better performance than other features [14].

## 5. Speaker Modeling

The next step after feature extraction is to generate patterns models for feature matching. In the training or recognition mode, speech models are built using the specific voice features extracted from the current speech samples. In the recognition mode, the speech model is used to compare with the current samples for identification or verification purposes. Three main types of modeling techniques are available, namely: template matching, stochastic modeling, neural networks. Various concepts were introduced under these techniques such as pattern matching (Dynamic Time Warping) which does direct template matching between training and testing subject. However, direct template matching is time consuming when the number of feature vectors increase. Clustering is a method to reduce the number of feature vectors by using a codebook to represent centres of the feature vectors (Vector Quantization). The LBG (Linde, Buzo and Gray) algorithm [15] and the k-means algorithm are some of the most well known algorithms for Vector Quantization (VQ). Other methods proposed includes neural networks and also stochastic models that use probability distribution such as Hidden Markov Model (HMM) and the Gaussian Mixture Model (GMM). In this project, the training models are generated using the Vector Quantization-LBG method. The speech feature coefficients are passed into the function to generate the codebook. The rationale for choosing the VQ-LBG method is the ease of implementation and comparable performance to other methods. The square Euclidean distance measurement for speech similarity measure will be used for testing.